\documentclass[floatfix,aps,prd,twocolumn,showpacs,10pt]{revtex4}
\usepackage{epsfig}
\usepackage{epsf}
\usepackage{amsmath}
\usepackage{amssymb}
\usepackage{graphicx}

\begin{document}



\title{CMB Constraint on Radion Evolution in the Brane World Scenario}

\author{K. C. Chan} \email{kcc274@nyu.edu}
\affiliation{ \label{affia1}	Department of Physics and Institute of Theoretical Physics, the Chinese University of Hong Kong, Hong Kong SAR, People's Republic of China}
\affiliation{Department of Physics, New York University, 4 Washington Place, New York 10003, USA} 
\author{M.-C. Chu} \email{mcchu@phy.cuhk.edu.hk}
\affiliation{ \label{affia1}	Department of Physics and Institute of Theoretical Physics, the Chinese University of Hong Kong, Hong Kong SAR, People's Republic of China}

\date{\today}

\begin{abstract}
 

In many versions of brane model, the modulus field of extra dimensions, the radion, could have cosmological evolution, which induces variation of the Higgs vacuum expectation value,  $\langle H \rangle$, resulting in cosmological variation of the electron mass $m_e$. The formation of Cosmic Microwave Background (CMB) anisotropies is thus affected, causing changes both in the peaks positions and amplitudes in the CMB power spectra. Using the three-year Wilkinson Microwave Anisotropy Probe (WMAP) CMB data, with the Hubble parameter $H_0$ fixed to be the Hubble Space Telescope (HST) result, 72 km s$^{-1}$ Mpc$^{-1}$, we obtain a constraint on $\rho$, the ratio of the value of $\langle H \rangle$ at CMB recombination to its present value, to be [0.97, 1.02].


\end{abstract}


\maketitle

\section{Introduction}
\label{sect:Intro}
    Investigations in string/M theory suggest the existence of extra dimensions. In particular, we may live in a brane world in a bulk of dimensions greater than 4. In this brane world picture, the matter fields are confined on the brane which is a (1+3)-dimensional hypersurface embedded in the bulk, while gravity can propagate freely in the whole extended universe. For a review on extra dimensions and brane worlds see \textit{e.g.}~\cite{Csaki,Brax1}. By restricting the standard model particles on the brane, brane world models have evaded the strong constraints on the size of extra dimensions. Up till now Newtonion gravity is only tested to submillimeter scale \cite{Kapner}, and existence of extra dimensions below this scale remains possible. From the phenomenological point of view, extra dimensions open a new door to address the hierarchy problem, the large scale discrepancy between the Planck scale and the electroweak scale \cite{ADD}.  For example, in the Randall-Sundrum two-brane model \cite{RS}, one can obtain a large scale discrepancy by fine-tuning the interbrane distance, the radion; thus some mechanisms have to be invoked to stabilize the radion \cite{Goldberger, Setare}.

      There are many moduli fields in string theory and the radion is an example. The moduli fields have to be stabilized so as not to upset the limits from long range force experiments and tests of general relativity. Other than invoking stabilization potentials, there could also be some attractor mechanism that drives the theory towards general relativity \cite{Brax03}. The radion may also act as a chameleon field and acquire mass by self-interactions to avoid the most restrictive of current bounds \cite{Khoury04, Mota}. If these scenarios hold, it is possible that the radion still evolves after the epoch of Big Bang Nucleosynthesis (BBN).

     Studies of brane world models can be divided into two main streams. On one hand, more complex models are studied in order to strengthen the connection between string theories and brane models; on the other hand, efforts are made to confront brane models with observational data \cite{Deruelle}.  In the brane world scenario, evolving radion induces variation of Higgs VEV, and therefore constraining the evolution of the radion implies limiting the evolution of Higgs VEV.  In Ref.~\cite{BaojiuBrane,BaojiuUED}, the constraints on the evolution of the radion are obtained from BBN and the best fit to the observational abundances of D, $^4$He and $^7$Li suggests a small variation in the radion during BBN compared to its present value. Molecular spectral lines can be used to determine $\mu \equiv m_p/ m_e$ \cite{Uzan}. Recently, Reinhold \textit{et al.} used quasar H$_2$ spectral lines to obtain $\Delta \mu /\mu = (2.4 \pm 0.6) \times 10^{-5}$ for a weighted fit \cite{Reinhold}. The constraints on $\mu$ can be interpreted as an indication that the Higgs VEV (and hence the radion) at redshift 2--3,  was different from its value today.  The Cosmic Microwave Background (CMB) anisotropies are measured with high precision by Wilkinson Microwave Anisotropy Probe (WMAP) \cite{WMAP3} and other observations. Thus it is timely to study the effects of the existence of extra dimensions on CMB and  fill the ``redshift gap'' between BBN and quasar constraints using data from CMB. In this article we shall use the three-year WMAP data to constrain the evolution of the radion.

    Within the brane world scenario, an extra term, called the dark radiation term, appears in the modified Friedmann equations \cite{ModifiedFried}, and it causes great difficulty in calculating the Sachs-Wolfe Effect \cite{Langlois01,Barrow,Koyama03,Rhodes03}.  Nonetheless, in this work we will constrain the radion evolution through the variation of constants caused by the variation of Higgs VEV ignoring the dark radiation term. CMB has been used to constrain the Higgs VEV using the pre-WMAP data in \cite{Yoo}. However it is desirable to get an improved bound using the high precision three-year WMAP data, including the polarization power spectrum. We also demonstrate that the constraints can be tightened substantially if we make use of HST measurement of $H_0$.

     This rest of the paper is organized as follows. In Section~\ref{sec:DR}, we carry out the conventional dimensional reduction to get the radion dependence of various constants. We shall see that only the Higgs VEV is expected to vary in the context of brane world models. We then describe the effect of variation of Higgs VEV on the CMB power spectra in Section~\ref{sec:effectCMB}. The numerical constraints on the radion (or Higgs VEV) are presented in Section~\ref{sec:NConstraint}. We summarize in Section~\ref{sec:Conclusion}.

\section{Dimensional reduction and low energy effective actions}
\label{sec:DR}
  In this section we will start from the higher dimensional action and carry out the dimensional reduction to derive the dependence of the constants on the radion. There are many works on effective actions in extra dimensions, \textit{e.g.} \cite{Brax03, Carroll, Sundrum}, and the formalism here follows closely that in Ref. \cite{Mazumdar,BaojiuBrane}. We shall dimensionally reduce the higher dimensional gravitational action to obtain its low energy effective action first. The higher dimensional action reads
\begin{equation}
\label{EqAction}
S=  \int d^{4+n}X \sqrt{-G}\left[  \left( \frac{1}{2 \kappa_{4+n}^2 } {}^{(4+n)}R\right) + \mathcal{L}_m \right],
\end{equation}
where $n$ denotes the number of compact extra dimensions, and $X^A$ represents the bulk spacetime coordinate, $A=0, 1, \dots, 3+n$. The higher dimensional Ricci scalar is denoted by ${}^{(4+n)}R$, and $G$ is the determinant of the full spacetime metric $G_{AB}$. $\mathcal{L}_m$ is the matter field Lagrangian density, which may include scalar fields, vector boson fields and Dirac fermion fields.

     To proceed we shall take the metric ansatz
\begin{equation}
\label{EqMetric}
ds^2= G_{AB}dX^A dX^B= g_{\mu \nu}(x)dx^{\mu} dx^{\nu} + h_{i j}(x)dy^i dy^j.
\end{equation}    
In this metric the Greek indices run over 0, 1, 2 and 3, while the Latin indices $i$ and  $j$ run over the extra dimensions from 4 to $3+n$. We are only interested in the zero-mode of the Kaluza Klein expansion, and so the extra dimensional metric $h_{ij}(x)$ does not depend on the extra dimension coordinate. Furthermore, we have assumed that the metric in  Eq.~\ref{EqMetric} is block-diagonal because the vector-like conection $G^{i}_{\mu}$ vanishes for zero-mode \cite{Mazumdar}.  The extra dimensions are compactified on an orbifold and the dimensionless coordinate $y^i$ assumes values in the interval [0,1]. 
Expressing ${}^{(4+n)}R$ in the terms of the 4D Ricci scalar ${}^{(4)}R$, $S$ can be written as 
\begin{eqnarray}
 S &  = &  \int d^{4}x d^n y \sqrt{-g}  \sqrt{h} \left[  \frac{1}{2 \kappa_{4}^2 \mathcal{V}_0 }   \left( {}^{(4)}R + \frac{1}{4}g^{\rho \sigma} \partial_{\rho} h^{i j} \partial_{\sigma} h_{i j}   \right.  \right.   
\nonumber \\
   &  +  &     \left.  \left.    \frac{1}{4}g^{\rho \sigma} h^{i j} \partial_{\rho}h_{i j} h^{k l} \partial_{\sigma}h_{k l} 
 \right) + \mathcal{L}_m \right],
\end{eqnarray}
where we have defined 
\begin{equation}
 \frac{1}{ \kappa_{4}^2} \equiv  \frac{ \mathcal{V}_{0} }{ \kappa_{4+n}^2 },
\end{equation}
with $\mathcal{V}_0$ being the volume of the extra dimensions today.
We shall work in the Einstein frame with pure Ricci scalar in the gravitational action. To do so we will apply the conformal transformation
\begin{equation}
\label{Eq:Conformal}
\tilde{g}_{\mu \nu} =e^{-2 \theta } g_{\mu \nu}
\end{equation}
with 
\begin{equation}
e^{-2 \theta} = \frac{\sqrt{h}}{\mathcal{V}_0}.
\end{equation}
For concreteness, we further assume that the extra dimensional manifold  is homogeneous and isotropic. Hence the extra dimension metric takes the simple form:
\begin{equation}
\label{Eq:EDMetric}
\mathrm{diag}(b^2,b^2,\ldots, b^2).
\end{equation}
   After the transformation Eq.~\ref{Eq:Conformal} and using the ansatz Eq.~\ref{Eq:EDMetric}, the action $S$ reduces to 
\begin{eqnarray}
S &  = &   \int d^{4}x  \sqrt{-\tilde{g}}  \left[  \frac{1}{2 \kappa_{4}^2 } {}^{(4)}R -   \frac{n(n+2)}{4 \kappa_4^2} \frac{1}{b^2}  \tilde{g}^{\mu \nu}  \partial_{\mu} b  \partial_{\nu}b  \right.   
 \nonumber   \\  
  & + &  \left.   e^{4 \theta}\mathcal{L}_m \right]. 
\end{eqnarray}
To make the scalar field canonical, we define a new scalar field, the radion $\sigma$, as\begin{equation}
\sigma \equiv \frac{1}{\kappa_4} \sqrt{\frac{n+2}{2n}} \ln \frac{b^n}{\mathcal{V}_0}. 
\end{equation}
Finally the effective action reads
\begin{equation}
S= \int d^{4}x  \sqrt{-\tilde{g}}  \left[  \frac{1}{2 \kappa_{4}^2 }\left( {}^{(4)}R -   \frac{1}{2} \tilde{g}^{\mu \nu}  \partial_{\mu} \sigma  \partial_{\nu}\sigma  \right)   + e^{4 \theta}  \mathcal{L}_m \right]. 
\end{equation}

   We now study the matter sector in more details. Although the matter fields only live on the brane, their actions are still affected by the existence of extra dimensions because of the conformal transformation Eq.~\ref{Eq:Conformal}.  Let us begin with the minimally coupled scalar field 
\begin{equation}    
S_{\rm Scalar} = \int d^4 x \sqrt{-g}\left(   - \frac{1}{2}g^{\mu \nu} \partial_{\mu}\phi \partial_{\nu}\phi - V(\phi) \right) .
\end{equation}
After the transformation Eq.~\ref{Eq:Conformal}, we have
\begin{eqnarray}
S_{\rm Scalar} & = & \int d^4 x \sqrt{-\tilde{g}} \left[-\frac{1}{2}  \tilde{g}^{\mu \nu}\partial_{\mu }\Phi  \partial_{\nu}\Phi    
\right.  \nonumber    \\
 & - &   \left.  \exp\left(- 2 \kappa_4 \sigma  \sqrt{\frac{2n}{n+2}} \right)   V(\phi)     \right],
\end{eqnarray}
where we have defined 
\begin{equation}
\Phi= \exp {\left(-\kappa_4 \sigma  \sqrt{\frac{n}{2(n+2)}}\right)} \phi,
\end{equation}
to make the kinetic term canonical.

If $V(\phi)$ is taken to be a simple renormalizable potential of the form
\begin{equation} 
V(\phi)= \frac{1}{2}\mu^2 \phi ^2 + \frac{1}{4}\nu \phi ^4,
\end{equation}
we then have
\begin{equation}
V(\Phi)= \frac{1}{2}\exp\left(- \kappa_4 \sigma \sqrt{\frac{2n}{n+2}} \right) \mu^2 \Phi^2  +\frac{1}{4}\nu \Phi^4.
\end{equation} 
We see that the mass of the scalar field is dependent on the radion in the Einstein frame for both signs of $\mu^2$. One of the most important scalar fields in particle physics is the Higgs field. Masses of the fermions and quarks can be generated by their Yukawa couplings to the Higgs field. In this mechanism, the fermion masses are proportional to the Higgs VEV, $\langle H \rangle$. Thus we get the radion dependence of the fermion mass
\begin{equation}
\label{eq:HiggsMass}
m \propto  \langle H \rangle   \propto \exp\left(- \frac{\kappa_4}{2} \sigma \sqrt{\frac{2n}{n+2}}  \right).
\end{equation}

       For the gauge field, the action is given by
\begin{equation}
S_{\rm Gauge} =- \frac{1}{4g_{*}^2} \int d^4 x \sqrt{-g}g^{\mu \rho}g^{\nu \sigma}   F_{\mu \nu} F_{\rho \sigma}, 
\end{equation}
where $F_{\mu \nu}$ is the gauge-invariant field strength tensor. However the action is invariant under the conformal transformation Eq.~\ref{Eq:Conformal}, and so the effective 4D coupling constant is the same as the higher dimensional $g_*^2$.

Similar techniques can be applied to the Dirac field $\psi$ with mass $\hat{m}$ \cite{Fujii}:
\begin{equation}    
S_{\rm Dirac}= \int d^4 x \sqrt{-g}\left(  i \bar{\psi}e^{i \mu} \gamma_{i} D_{\mu} \psi - \hat{m} \bar{\psi}\psi \right) ,
\end{equation}
where $e^{i \mu}$ is the vierbein and $D_{\mu}$ is the covariant derivative. 
Applying Eq.~\ref{Eq:Conformal} and redefining the field $\psi$ as
\begin{equation} 
\Psi = \exp \left( -\frac{3\kappa_4}{4}\sigma  \sqrt{ \frac{2n}{n+2}} \right) \psi,
\end{equation}
we get the canonical action
\begin{eqnarray}
S_{\rm Dirac} & = & \int d^4 x \sqrt{-\tilde{g}}\left[ i \bar{\Psi}  \tilde{e}^{i \mu}   \gamma_i \tilde{D}_{\mu} \Psi     \right.   
  \nonumber  \\   
 & - &  \left.           \hat{m} \exp\left(- \frac{\kappa_4}{2} \sigma  \sqrt{\frac{2n}{n+2}} \right) \bar{\Psi} \Psi   \right].
\end{eqnarray}
We can read out the radion dependence of the fermion mass, which is the same as Eq.~\ref{eq:HiggsMass}.

    Thus, for the brane world models in the Einstein frame, only the fermion masses, among other fundamental constants, acquire radion dependence and are expected to vary. Or, in the framework of Standard Model, we may say that only  $\langle H \rangle$ is radion-dependent while the Yukawa coupling is constant. 

\section{Numerical Constraints on the Evolution of the radion}
\label{sec:constraint}

       In the last section, we see that only $\langle H \rangle$ is expected to vary in the brane world scenario. Here we first discuss the effects of variation of $\langle H \rangle$ on the CMB power spectra, and then we present the numerical constraints on $\langle H \rangle$ and the radion using the the three-year WMAP data.

\subsection{Effects of variation of Higgs VEV on CMB power spectra}
\label{sec:effectCMB}

    The variation of $\langle H \rangle$  induces changes in the Fermi constant $G_F$, the quark masses, the nucleon binding energy and the electron mass $m_e$ \cite{Dixit}. The variation of $G_F$ is not relevant in the epoch of CMB recombination. We can also ignore the changes in strong nuclear force and nuclear binding energy caused by the variation of quark mass. Since the baryon mass is dominated by the QCD scale parameter $\Lambda_{\rm QCD}$, and the quark masses contribute at only a few percents level, we shall neglect the influence of  $\langle H \rangle$ on the baryon mass. Thus the variation in $\langle H \rangle$  boils down to variation of $m_e$ \cite{Kujat}, which modifies the recombination history.

    The variation of $m_e$ enters CMB through the binding energies of  hydrogen, H$_{\rm I} $, and helium, He$_{\rm II}$, Thomson cross-section $\sigma_{T}$, and the recombination coefficients $\alpha$ and the two-photon decay rates of hydrogen and helium. In the Appendix, we list the evolution equations and modifications to take care of variations of $m_e$. Here we only sketch the modifications to be made. The binding energies scale as $m_e$, $\sigma_{T}$ is proportional to $m_e^{-2}$, and the two-photon decay rates vary as $m_e$ \cite{BT}. For $\alpha$, one can derive a differential equation relating the matter temperature $T_M$ and $m_e$ \cite{Kujat}, through which the dependence of $\alpha$ on $m_e$ can be deduced using the empirical fitting of $\alpha$ as a function of $T_M$ in the literature.

    Among these effects caused by the variation of $m_e$, the change in the binding energy of hydrogen is most significant on the power spectra; less important is the effect of Thomson cross-section, while the effects of the recombination coefficients and two-photon rates are small.

    For convenience, we define 
\begin{equation}
\rho \equiv \frac{\langle  H \rangle  _{\rm CMB}}{\langle H \rangle_0},
\end{equation}
where $\langle H \rangle_{\rm CMB}$ and $\langle H \rangle_0$ denote the Higgs VEV in the era of CMB recombination and today respectively.

\begin{figure}
	\centering
	\includegraphics[width=0.8\linewidth] {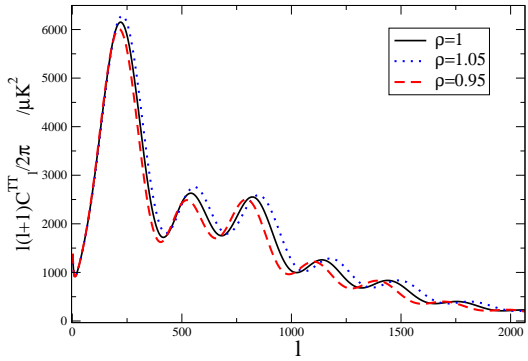}
        \caption{The CMB temperature power spectrum with $\rho$=1, 1.05 and 0.95 respectively. Other cosmological parameters assume the standard values. }         
	\label{fig:TTspectrum}
\end{figure}

     Shown in Fig.~\ref{fig:TTspectrum} is the CMB temperature power spectrum with $\rho$ being  1, 1.05 and 0.95 respectively with the other parameters being the standard ones. 
We see that both the positions and amplitudes of the peaks change for $\rho=1.05$ and 0.95 in comparison with $\rho=1$. For $\rho=1.05$ (0.95), because the binding energy $\propto m_e$ and it dictates the epoch of CMB recombination, CMB recombination takes place at an earlier (later) time. Therefore the sound horizon is smaller (larger), and the distance to the last scattering surface is larger (smaller); the peaks shift to the larger (smaller) $l$ scales. The change in amplitudes is due to early Integrated Sachs--Wolfe (ISW) effect and damping effect. For $\rho=1.05$ (0.95), the residual radiation is higher (lower) right after recombination, which gives an enhanced (reduced) early ISW effect and hence the first peak is boosted (diminished). Early recombination also means greater Hubble rate and narrower visibility function at decoupling time, and thus the damping is reduced and the power in high $l$ scales is enhanced.

\begin{figure}
	\centering
	\includegraphics[width=0.8\linewidth] {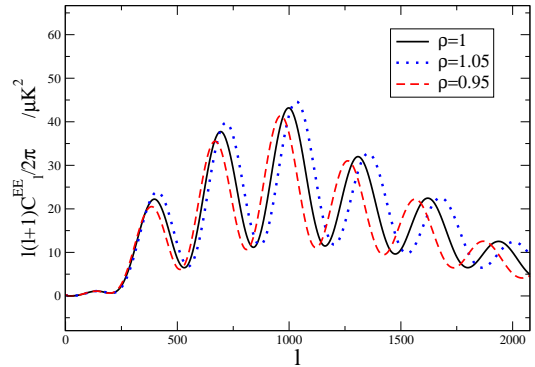}
        \caption{The CMB $E$-polarization power spectrum with $\rho$=1, 1.05 and 0.95 respectively.  }             
	\label{fig:EEspectrum}
\end{figure}

      Shown in Fig.~\ref{fig:EEspectrum} is the $E$-polarization power spectrum with $\rho$=1, 1.05 and 0.95 respectively. The main features of this power spectrum can be understood with the tight coupling semi-analytic formula for the polarization strength $\Theta_{Pl}$ of a particular $k$ mode \cite{Dodelson}:
\begin{equation}
\Theta_{Pl}(k) \simeq \frac{5 k \Theta_1 (k, \eta_{*})} {6 \dot{\tau} (\eta_{*})}  \frac{l^2 }{ [ k (\eta_{0}- \eta_{*})]^2} j_{l}(k (\eta_{0} - \eta_{*} ) ),
\end{equation} 
where $\eta_{0}$ and $\eta_{*}$ are the conformal times of today and the last-scattering surface respectively. $\Theta_{1}(k,\eta_{*})$ is the dipole of the photon distribution. Again, for larger $m_e$, recombination occurs at an earlier conformal time, and because the spherical Bessel function peaks at $l \sim k (\eta_{0} - \eta_{*} )$, the peaks shift to the larger $l$s. The power of the peaks is higher because the dipole, which is like the velocity of a fluid in the tight coupling limit, is larger at the earlier time.


     In the above discussions, we have assumed that $m_e$ is constant in the whole calculation.  Because the effect of $m_e$ on CMB is only important near the decoupling time, even if we assume that $m_e$ evolves, \textit{e.g.} as a linear function of the scale factor $a$, only the value of $m_e$ at the decoupling time matters. Thus CMB indeed gives the constraint on $m_e$ at $z \sim 1000$.

\subsection{Numerical constraints on the Higgs VEV and radion by CMB}
\label{sec:NConstraint}  

In this section, we constrain the range of $\langle H \rangle $  (or radion) using the three-year WMAP data \cite{WMAP3}. To do so, we make use of the Markov Chain Monte Carlo (MCMC) method implemented by the engine CosmoMC \cite{Cosmomc}, which searches for the maximum of the likelihood function. The theoretical CMB spectra are calculated by the Boltzmann code CMBFAST \cite{CMBFAST}. We vary the following set of parameters: the Hubble parameter, $ H_0$, the baryon density, $\omega_b = \Omega_b h^2$ ($h=H_0/100$ km s$^{-1}$ Mpc$^{-1}$), the cold dark matter density, $\omega_c =\Omega_c h^2 $, the reionization redshift, $z_{\rm re} $, the primordial fluctuation amplitude, $A_s$, the spectral index, $n_s$, and $\rho$. Flatness of the universe is assumed in all the calculations.

\begin{figure}
	\centering
	\includegraphics[width=0.8\linewidth] {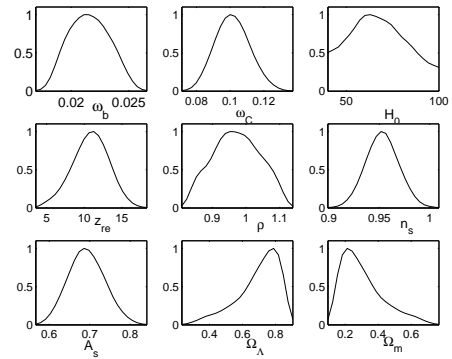}
        \caption{The marginal distributions of the free parameters, constrained by the three-year WMAP data. In addition to the free parameters $\omega_b \equiv \Omega_b h^2$, $\omega_c=\Omega_c h^2 $, $H_0$, $z_{re}$, $n_s$, $A_s$ and  $\rho$, shown also are the derived distributions of the density parameter of matter ($\Omega_m$) and cosmological constant ($\Omega_\Lambda$). The Hubble parameter is allowed to vary in the range from 40 to 100 km s$^{-1}$ Mpc$^{-1}$ (we will suppress this unit afterwards). Here and thereafter, the maxima of the distributions are arbitrarily normalized to 1.}             
	\label{fig:BraneH_dis}
\end{figure}

\begin{figure}
	\centering
	\includegraphics[width=0.8\linewidth] {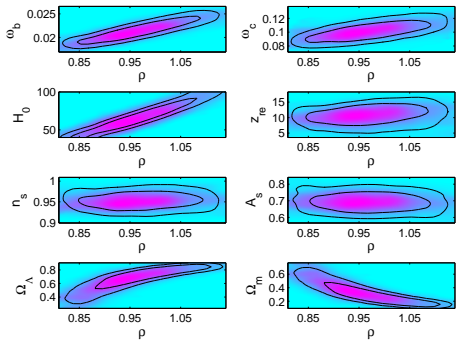}
        \caption{The contour marginal distributions of cosmological parameters plotted against $\rho$.  Among the free parameters, $H_0$ is the most strongly degenerate with $\rho$.  }             
	\label{fig:BraneH_degen}
\end{figure}

\begin{figure}
	\centering
	\includegraphics[width=0.8\linewidth] {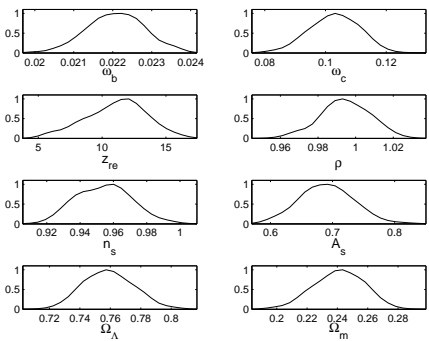}
        \caption{Same as Fig.~\ref{fig:BraneH_dis}, but with $H_0$ fixed to be 72. The spread of $\rho$ is much smaller than that in Fig.~\ref{fig:BraneH_dis}. }             
	\label{fig:BraneH0fix}
\end{figure}

 The marginal distributions of the parameters are shown in Fig.~\ref{fig:BraneH_dis}.
In particular, the 95\% confidence interval (C.~I.) for $\rho$ is $[0.85, 1.11]$. The constraint is relatively weak. Most of the other standard cosmological parameters are well constrained in the usual ranges, except the Hubble parameter. Although the mean of $H_0$ being 67.6 is still close to the ``canonical'' value 72, its distribution spreads wide. This suggests that $H_0$ may be degenerate with $\rho$.  To see the degeneracy between $\rho$ and other parameters explicitly, we display the contour marginal distributions of various cosmological parameters plotted versus $\rho$ in Fig.~\ref{fig:BraneH_degen}.   
Among the parameters, $\omega_b$ and $\omega_c$ show slight degeneracy with $\rho$, but $\rho$ is strongly degenerate with $H_0$. This suggests that we may obtain a tighter bound if we use the measurement of $H_0$ by the Hubble Space Telescope (HST) Key Project to break the degeneracy between $H_0$ and $\rho$. We carry out the MCMC run once again but with $H_0 =72$ \cite{HST}. Indeed, the bounds on $\rho$ are tightened substantially, as can be seen in Fig.~\ref{fig:BraneH0fix}. The 95\% C. I. now becomes [0.97, 1.02].

\begin{figure}
	\centering
	\includegraphics[width=0.8\linewidth] {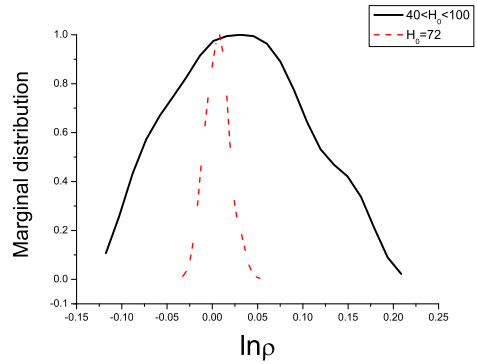}
        \caption{The derived marginal distributions of $ \ln \rho = - \frac{\kappa_4}{2}\sqrt{ \frac{n+2}{2n}}   ( \sigma_{\rm CMB} - \sigma_0 )   $ for $40<H_0 <100$ (solid line) and $H_0 =72$ (dashed line) respectively. Note that the latter has much smaller spread. }             
	\label{fig:radion_dis}
\end{figure}

Using  Eq.~\ref{eq:HiggsMass}, we get the change of the radion
\begin{equation}
\label{eq:sigma_rho}
\sigma_{\rm CMB} - \sigma_0  = - \frac{2}{\kappa_4} \sqrt{\frac{2n}{n+2}} \ln \rho ,
\end{equation}
which is proportional to $\ln \rho$. In Fig.~\ref{fig:radion_dis}, we show the derived distribution of $\ln \rho$.    
One sees again that the spread of the distribution of $( \sigma_{\rm CMB} - \sigma_0) $ for $H_0$ being allowed to vary  from 40 to 100 is much larger than that for fixed $H_0=72$, and their 95\% C. I. are [-0.094, 0.17] and [-0.019, 0.034] respectively.

     The constraint on Higgs VEV by CMB is first deduced in \cite{Yoo} using pre-WMAP data sets. There is no marked difference between their constraint $0.92< \rho <1.13$ and ours with $H_0$ being allowed to vary in the interval $40<H_0<100$. The small difference may be due to the use of different data sets and/or statistical methods. The data set we used is the recent three-year WMAP data with the polarization spectrum \cite{WMAP3} while the pre-WMAP temperature power spectrum is used in \cite{Yoo}. They restricted $H_0$ in the interval [50, 80] and more importantly $\Omega_M$ in the interval [0.3, 0.4], which is a small portion of the distribution in Fig.~\ref{fig:radion_dis}. We also allow for two more free parameters, $z_{\rm re}$ and $A_s$ in our MCMC runs. That's why we get bounds comparable to theirs although we use more accurate data set.  Moreover, we show that the bounds on $\rho$ can be tightened substantially if the HST measurement of $H_0$ is used.


     Since the upcoming Planck satellite mission is going to measure the temperature power spectrum to as high as $l \sim  2500$ and the $E$-polarization spectrum to $l \sim 1500$, we expect that there will be tremendous improvement in the constraint on $\rho$. We can forecast the improvement that Planck will bring quantitatively using the Fisher matrix, which has been widely used to predict the expected uncertainties in future experiments (see \textit{e.g.} \cite{Eisenstein}). Under the assumption of Gaussian perturbations and Gaussian noise, the Fisher matrix takes the form
\begin{equation}
F_{i j} = \sum_{l} \sum_{X,Y} \frac{\partial C _{Xl} }{\partial p_i}  (\mathrm{Cov}_{lXY})^{-1} \frac{\partial C_{Yl} }{\partial p_j},
\end{equation}
where $p_i$ is the $i$th free parameter and $C_{X l}$ is the $l$th multipole of the observed spectrum of type $X$, which can be the temperature, temperature-polarization and E-polarization spectra. The experimental precision is encoded in the covariant matrix $\mathrm{Cov}_{lXY}$. With the expected Planck power spectra, the constraint on $\rho$ is tightened by a factor of 7 when $H_0$ is allowed to vary; the current constraint with $H_0$ fixed will be tightened by a factor of 5.

    We compare the present CMB constraints with other constraints in Table \ref{tab:me_constraint_table}. We note that the BBN constraint and the quasar absorption line constraint are of opposite signs. Even if both constraints are found to be correct, that does not immediately rule out the evolution of the radion since the radion may stabilize to its present value in an oscillatory manner. If this is the case, we may find that the radion takes a value very close to 1 near the epoch of CMB recombination. Unfortunately, the current CMB data is not discriminating enough. With the forthcoming Planck data, the CMB constraint will be tightened substantially and may tell us if this interesting scenario holds or not.

\begin{table}[htb!]
\caption{The present constraints on the variation of $\langle H \rangle $ at various redshifts. The CMB constraint fills the gap in the ``redshift ladder'' in between BBN and quasar. The future Planck data will tighten the CMB constraint ( so that it is compatible with other constraints.) } 
\label{tab:me_constraint_table}
\begin{tabular}{|l |    p{0.25\linewidth} |p{0.5\linewidth} |}
\hline
 Redshift $z$  &  Observations   & 95 \% C.~I. for $(\langle H \rangle_z -  \langle H \rangle_0  ) / \langle H \rangle_0 $, where $\langle H \rangle_z$ denotes the value of Higgs VEV at redshift $z$     \\

\hline

  $ 10^{10} $   &       BBN \cite{BaojiuBrane}      &   [0.00,0.04]      \\
\hline

 1000 &   CMB   & $H_0$ free: [-0.15,0.11]; $H_0$=72: [-0.03, 0.02]    \\
\hline
2-3    &  Quasar absorption lines \cite{Reinhold}    &  Weighted fit: $- 0.9 \times [1.8 \times 10 ^{-5}, 3.0 \times 10^{-5}]   $  \footnote{The prefactor 0.9 is due to our assumption that 10\% of the proton mass is contributed by the rest mass of the quarks.}  \\
\hline

\end{tabular}
\end{table}

\section{Conclusion}
\label{sec:Conclusion}  

In the brane world scenario, the evolution of radion induces variation of $\langle H \rangle $. Among its consequences, we expect only the variation of $m_e$ to be relevant in the epoch of CMB recombination. Variation of $m_e$ changes the time of CMB recombination, causing changes in both the peak positions and amplitudes of the CMB power spectra. Thus we can constrain the radion evolution via limiting the variation of $\langle H \rangle$ using the CMB data. With the three-year WMAP data and $H_0$ as a free parameter, we obtain the 95\% C. I. for $\rho$ to be [0.85, 1.11]; when we fix $H_0$ to be the HST result 72 km s$^{-1}$ Mpc$^{-1}$ the constraint for $\rho$ is tightened to be [0.97, 1.02]. In terms of $\ln \rho$, which is proportional to the change in radion $\sigma$ by Eq.~\ref{eq:sigma_rho}, the corresponding 95\% C. I.'s are [-0.094, 0.17] and [-0.019, 0.034] respectively. Although the current CMB data is not discriminating enough, the upcoming Planck data should tighten the bounds by a factor of 5 or so.

\section*{Appendix}
       
  In this appendix we list the evolution equations for calculating the ionization history with the effect of modified $m_e$ taken into account. The detailed evolution of the ionization fraction $x_e$ can be modeled by the simple Peebles recombination \cite{Peebles}, which is basically a two-level approximation, or using RECFAST \cite{RECFAST}, the evolution equations of which have been checked against more detailed multi-level calculations. We use a modified RECFAST which allows for variation of $m_e$ to evaluate the evolution of $x_e$, as in Ref.~\cite{Ichikawa}. The ionization history can be modeled by evolving the ionization fraction of H, $x_{p}$, ionization fraction of $\rm He_I $, $x_{ \rm He_{II} }$ and the matter temperature $T_{M}$ (see \cite{RECFAST} for details):
\begin{eqnarray}
 \frac{dx_{p}}{dz} &  = &  \frac{ C_{\rm H}}{ H(z)(1+z)} \left[ x_e x_p n_{\rm H} \alpha{\rm H }  \right.  \nonumber  \\   &   - &   \left.   \beta_{\rm H} (1- x_p ) \exp \left( - \frac{h \nu_{\rm H 2s} }{k T_{M}} \right)  \right]
\end{eqnarray} 
with
\begin{equation}
C_{\rm H} = \frac{1+ K_{\rm H} \Lambda_{\rm H} n_{\rm H} (1-x_p)  }{ 1+ K_{\rm H} (\Lambda_{\rm H}+\beta_{\rm H} ) n_{\rm H} ( 1- x_{p}) } , 
\end{equation}
\begin{eqnarray}
\frac{ d x_{\rm He_{II} } } {dz} &  =  &
\frac{C_{\rm He_{II} } }{ H(z)(1+z) }  
 \left[    x_{\rm He_{II}}  x_{e} n_{\rm H} \alpha_{\rm He_{I}}         
 \right.  \nonumber  \\  
& - &   \left.
  \beta_{\rm He_{I}}   (  f_{\rm{He}} - x_{\rm{He II}   })   
 \exp \left(   -\frac{ h \nu_{\rm He_{ I  2s}  }} {kT_{M}} 
  \right) \right] , 
\end{eqnarray}
with
\begin{equation}
C_{\rm He_{II} }  =   \frac{1 + K_{\rm{He_I}} \Lambda_{\rm{He}} n_{\rm H}
 (f_{\rm He}-x_{\rm He_{II} })    \exp \left(  -\frac{h \nu_{ \rm He_{ I } 2p- 2s  }}{k T_{M}} \right) } 
 { 1 + K_{\rm He_I  } (\Lambda_{\rm He} + \beta_{\rm He_I} )  n_{\rm H}
 ( f_{\rm He} - x_{ \rm He_{II} }  )   
 \exp \left(  - \frac{ h \nu_{ \rm He_{ I } 2p- 2s }} { k T_{M}}  \right)   } , 
\end{equation}
and
\begin{equation}
\frac{dT_{M}}{dz}= \frac{8 \sigma_{T} a_{R} T^4_{R}}{3 m_e c H(z)(1+z) } \frac{x_e}{1+ f_{\rm He} +x_e} (T_M -T_R) + \frac{ 2 T_M}{1+z} .
\end{equation}

  The meanings of the symbols are in order. In these equations, $z$ is the redshift and $H(z)$ is the Hubble expansion rate at $z$, and $n_H$ is the total hydrogen number density. $T_R$ denotes the radiation temperature $T_0(1+z)$.

     $\nu_{\rm H 2s}=c$/121.5682 nm is the Lyman $\alpha$ frequency. For helium, $ \nu_{\rm He_{ I  2s}} = c$/60.1404 nm, and  $\nu_{ \rm He_{ I } 2p- 2s }$ denotes $c/58.4334$ nm $-$  $c/60.1404$ nm, the frequency difference between He$_{\rm I}$ $2^1 p$ and He$_{\rm I}$ $2^1 s$. $K_{\rm H} \equiv ( 121.5682 \mbox{ nm})^3 /[ 8 \pi H(z)] $ and $K_{\rm He_{I}} \equiv ( 58.4334 \mbox{ nm})^3 /[ 8 \pi H(z)] $ give the amount of cosmological redshiftings of  $2^{1}p$ to $1^{1}s$ photons. These frequency $\nu$ dependent quantities  need to be modified as $\nu \propto m_e $.

    $\Lambda_{\rm H}$ and $\Lambda_{\rm He}$ denote the two-photon decay rates from $2^{1}s$ to $1^{1}s$ for H and He respectively, and they scale as $m_e$ \cite{BT}.  

     The Thomson cross-section $\sigma_{T}$ scales as $m_e^{-2}$ as $\sigma_{T}=8 \pi e^4/(3 m_e^2 c^4)$.

     $\alpha{\rm H }$ ($\alpha_{\rm He_{\rm I}}$) is the recombination coefficient of H$_{\rm I}$ (He$_{\rm II}$), and the photo-ionization $\beta_{\rm H}$ ($\beta_{\rm He_{\rm II} }$)  is related to it by 
\begin{equation}
\beta =  \alpha  \left(  \frac{2 \pi m_e k T_M }{h^2} \right)^{3/2}  \exp\left( \frac{ -h \nu_{\rm 2s} }{k T_M} \right).  
\end{equation}
The recombination coefficient $\alpha$ can be expressed as \cite{Kaplinghat}
\begin{equation}
\alpha  = \sum_{n,l}^{*} 8 \pi (2l+1) \left(  \frac{k T_M }{2 \pi m_e} \right)^{\frac{3}{2}}  \exp \left( \frac{B_n}{kT_M}  \right) \int_{\frac{B_n}{kT_M}}^{\infty}  \frac{\sigma_{nl} y^2  dy }{e^y -1 },  
\end{equation}
where $B_n$ is the binding energy of the $n$th state and $\sigma_{nl}$ is the ionization cross-section. The asterisk indicates that the sum should be regulated. Using the fact that $B_n$ scales with $m_e$ and the ionization cross-section scales as $m_e^{-2}$ \cite{Uzan}, we can derive \cite{Kujat}
\begin{equation} 
\label{eq:m_eT_M}
\frac{   \partial \alpha}{ \partial m_e} = - \frac{1}{m_e} \left( 2 \alpha + T_M \frac{\partial  \alpha }{\partial T_M}  \right)      .  
\end{equation}
In the literature  $\alpha$ is usually parametrized as a function of $T_M$ in the fitting formula. The $m_e$ dependence of $\alpha$ can be extracted from the fitting formula using Eq.~\ref{eq:m_eT_M} and hence also the $m_e$ dependence of $\beta$.

\appendix

\begin{acknowledgments}
We are grateful for B. Li for useful discussions. We also thank the ITSC of the Chinese University of Hong Kong for providing its clusters for computations. This work is partially supported by a grant from the Research Grant Council  of the Hong Kong Special Administrative Region, China (Project No. 400707).    
\end{acknowledgments}


\end{document}